\documentclass[12pt,a4paper]{article}
\usepackage{amssymb,cite,graphics}
\newcommand\blank[1]{}

\newcommand{\resection}[1]{\setcounter{equation}{0}\section{#1}}

\begin{document}
\begin{titlepage}
\vskip 0.5cm

\vskip 1.5cm
\begin{center}
{\Large{\bf
 Second order quantum corrections to the classical reflection
 factor of the sinh-Gordon model
}}
\end{center}
\vskip 0.8cm
\centerline{A. Chenaghlou
\footnote{{\tt a.chenaghlou@sut.ac.ir}}}
\vskip 0.9cm
\centerline{\sl\small Physics Department,  Faculty of Science,
 Sahand University of Technology,  \,}
\centerline{\sl\small
 P O Box 51335-1996,  Tabriz,  Iran\,}
\vskip 1.25cm
\begin{abstract}
\noindent
The sinh-Gordon model on a half-line with integrable boundary
conditions is considered in low order perturbation theory
developed in affine Toda field theory. The quantum corrections to
the classical reflection factor of the model are studied up to the
second order in the difference of the two boundary parameters and
to one loop order in the bulk coupling. It is noticed  that the
general form of the second order quantum corrections are
consistent with Ghoshal's formula.
\end{abstract}
\end{titlepage}
\setcounter{footnote}{0} 
\def\thefootnote{\fnsymbol{footnote}}

\resection{Introduction}

Affine Toda field theory \cite{AFZ,MOP} is one of the most
successful examples of the two-dimensional quantum field theory
since it can be solved exactly. This theory possesses remarkable
properties including classical and quantum integrability. An
interesting review of the recent developments in Affine Toda field
theory is presented in Ref.\cite{C1}. The classical affine Toda
field theories remain integrable in the presence of certain
boundary conditions restricting them to a half-line, or to an
interval \cite{CDRS,CDR,BCDR,FS,FK1,FK2}. Indeed,  Corrigan et.al
\cite{CDRS,CDR,BCDR} have classified the boundary conditions which
preserve classical integrability.
 However, quantum integrability are hardly explored in the presence of a boundary although
 there has been progress   for models based on $a_{n}^{(1)}$ class
 of theories
\cite{CDRS,Ga,DG,PB,CH}. The simplest of these (n=1) is the
sinh-Gordon model. In fact, this model is the only example in the
$ade$ series of affine Toda field theory which allows continuous
boundary parameters. Although the sinh-Gordon model has been
studied much more than other models in the integrable quantum
field theories there still remains much to be studied in
connection with the quantum integrability of the model with a
boundary. In particular, one of the interesting problems is to see
how the boundary parameters could relate to the quantum reflection
factor of the theory.

In 1993 Ghoshal and Zamolodchikov \cite{GZ} studied the behaviour
of the sine-Gordon model restricted to a half-line. They found the
soliton reflection factors in the model and discovered a two
parameter family of boundary conditions which preserved
integrability. Then, Ghoshal \cite{G} computed the reflection
factors of the soliton-antisoliton bound states (the breathers) of
the model. The pioneering ideas of Ghoshal and Zamolodchikov
motivated many researchers  to investigate integrable quantum field
theory with a boundary. Corrigan \cite{C} was the first to
discover that the lightest breather reflection factor of the
sine-Gordon model is identical to the reflection factor of the
sinh-Gordon model followed by an analytic continuation in the
coupling constant.

Recently, Corrigan and Delius \cite{CD} calculated the bound-state
spectrum of the sinh-Gordon model on a half-line in two different
ways, firstly by using a boundary bootstrap principal, and
secondly by quantizing the classical boundary breather states
using a WKB technique (for further discussion see Ref.\cite{C2}).
Comparing these two calculations provides a relationship between
the boundary parameters, the bulk coupling constant and the
parameters in Ghoshal's formula. They performed the calculations
in the special case where the bulk $Z_{2}$ symmetry is conserved
at the boundary, by requiring the two boundary parameters to be
equal. Then, Corrigan and Taormina \cite{CT} studied and extended
the idea of \cite{CD} in the intricate case where the two boundary
parameters are different and the bulk symmetry is violated.
 Their calculations lead to a determination of the relationship
 between the Lagrangian boundary parameters and the two parameters
 appearing in the reflection factor of the sinh-Gordon particle.
Moreover, their results clarify the weak-strong coupling duality
of the sinh-Gordon model with integrable boundary conditions.

The one loop quantum corrections to the classical reflection
factor of the sinh-Gordon model have been found \cite{C} when the
boundary parameters are equal. If the two boundary parameters are
different and the bulk $Z_{2}$ symmetry of the model is broken
then, the lowest energy static background configuration will not
be a trivial solution ($ \phi=0 $). In this intricate case, the
perturbation theory must be developed within a complicated
background. Meanwhile,  the corresponding calculations involve a
significantly more difficult propagator as well as intricate
coupling constants. In a recent paper \cite{CC} the quantum
reflection factor has been found in one loop order up to the first
order in the difference of the boundary parameters. The
calculations provide a further  verification of
Ghoshal's formula up to the first order. The quantum reflection factor up to the second
order in the difference of the boundary data is treated in this
article. The result could test more deeply the expression for the
sinh-Gordon model reflection factor given in Ref. \cite{G}.

\resection{The  sinh-Gordon model on the half-line}

The sinh-Gordon theory describes a real massive scalar quantum
field theory in 1+1 dimensions with exponential self-interaction.
  The Lagrangian density of the theory in the presence of a
  boundary is
\begin{equation}
\bar{\mathcal{L}}=\theta(-x) \mathcal{L} -\delta(x) \mathcal{B}
\end{equation}
where  the bulk Lagrangian density of the model  is
\begin{eqnarray}
\mathcal{L}&=&\frac{1}{2}\partial_{\mu}\phi\partial^{\mu}\phi -
V(\phi) \nonumber\\
&=&\frac{1}{2}\partial_{\mu}\phi\partial^{\mu}\phi
-\frac{2m^{2}}{\beta^{2}}\cosh( \sqrt{2} \beta  \phi),
\end{eqnarray}
and the boundary potential $\mathcal{B}$ has the following generic
form \cite{GZ}
\begin{equation}
\mathcal{B}=\frac{m}{\beta^{2}}\left(\sigma_{0}e^{-\frac{\beta}
{\sqrt{2}}\phi}+
\sigma_{1}e^{\frac{\beta}{\sqrt{2}}\phi} \right).
\end{equation}
Note, in the above expressions $m$ and $\beta$ are a mass scale
and a coupling constant of the theory and we have used
normalizations customary in affine Toda field theory. The two real
coefficients $\sigma_{0}$ and $\sigma_{1}$ are essentially free
which indicate \cite{CDR,FS} the extra parameters permitted at the
boundary $x=0$. Meanwhile, the boundary potential is required to
satisfy the following equation as a consequence of maintaining
integrability on the half-line:
\begin{equation}
\frac{\partial\phi}{\partial x}=-\frac{\partial\mathcal{B}}
{\partial\phi}
 \hspace{.25in}   \hbox{at}\,\,\,\,  x=0,
\end{equation}
or
\begin{equation}
\frac{\partial \phi}{\partial x}=-\frac{\sqrt{2}}{\beta} \left(
\sigma_{1} e^{\beta \phi/\sqrt{2}} - \sigma_{0} e^{- \beta
\phi/\sqrt{2}} \right) \hspace{.25in}      \hbox{at}      \,\,\,\,
x = 0,
\end{equation}
where we have rescaled the mass scale to unity in the integrable
boundary condition. It is also convenient to use an alternative
expression for the boundary parameters, i.e. $\sigma_{i}=\cos
a_{i}\pi$.

The sinh-Gordon model is integrable classically which implies
there are infinitely many mutually commuting,  independent
conserved charges $Q_{\pm s}$, where s is an arbitrary odd
integer. On the other hand, the model is integrable in the context
of quantum field theory  which means the S-matrix describing the
n-particles scattering factorises into a product of two-particles
scattering amplitudes. The S-matrix describing the elastic
scattering of two sinh-Gordon particles with relative rapidity
$\theta$ is conjectured to be given by  \cite{AFZ,FK,ZZ}
\begin{equation}
S(\theta)=-\frac{1}{(B)(2-B)}
\end{equation}
where we have used  the hyperbolic building  block notation
\begin{equation}
(x)=\frac{\sinh(\theta/2+ \frac{i\pi
x}{4})}{\sinh(\theta/2-\frac{i\pi
x}{4})},
\end{equation}
 and the quantity B is related to the coupling constant $\beta$ by
$B= \frac{2\beta^{2}}{4\pi+\beta^{2}}$.

 For the boundary
sinh-Gordon model, the most important physical quantity is the
boundary S-matrix or the reflection factor which describes the
reflection of a single particle from the boundary. Firstly,
Ghoshal and Zamolodchikov \cite{GZ} calculated the soliton
reflection factors for the sine-Gordon model by solving the
boundary Yang-Baxter equation and using general constraints
implementing unitarity and a form of crossing symmetry. Then,
Ghoshal \cite{G} calculated the soliton-antisoliton bound states
reflection factor by using  the boundary bootstrap equation.
Finally, following the idea discovered by Corrigan \cite{C}, the
quantum reflection factor for the lightest breather in the
sine-Gordon model is supposed to be identical with the quantum
reflection factor of the sinh-Gordon particle after doing analytic
continuation in the coupling constant to derive

\begin{equation}
K_{q}(\theta)=\frac{(1)(2-B/2)(1+B/2)}{(1-E(\sigma_{0},\sigma_{1},
\beta))
(1+E(\sigma_{0},\sigma_{1},\beta) )(1-F(\sigma_{0},\sigma_{1},\beta)
)(1+F(\sigma_{0},\sigma_{1},\beta) )}.
\end{equation}
When the bulk reflection symmetry is preserved which means
$\sigma_{0}=\sigma_{1}$, then one of two parameters $E$ or $F$
vanishes.

Recently, on the basis of a perturbative calculation, the form of
the $E$ and  $F$ was conjectured to be \cite{CC},
\begin{equation}\label{conjecture}
E=(a_0+a_1)(1-B/2), \qquad F=(a_0-a_1)(1-B/2).
\end{equation}
In fact,  Corrigan and Taormina \cite{CT} gave further evidence
for these formulae.  They computed the energy spectrum of the
quantized boundary states, firstly by using a bootstrap technique
and, secondly using a WKB approximation. Requiring that the two
methods agree with each other provides further support to the
above conjecture. Meanwhile, they did the non-perturbative
calculations in the significantly more difficult case when
 $\sigma_{0} \neq \sigma_{1}$.

\resection{Low order perturbation theory}

Since perturbative expansion will be developed around an
$x$-dependent static background field configuration, it is
possible to use standard Feynman rules in configuration space. We
will find that the theory needs three- and four-point couplings
which depend upon the $x$-dependent static background. In other
words, the interaction vertices in the Feynman diagrams will be
position dependent. In order to derive the three and four-point
couplings, we need to expand the bulk and boundary potentials in
terms of the coupling constant $\beta$ around the background
solution. We obtain the bulk three- and four-point couplings:

\begin{equation}
C_{bulk}^{(3)}=\frac{2\sqrt{2}}{3}
\beta\sinh(\sqrt{2}\beta \phi_{0})
\end{equation}

\begin{equation}
C_{bulk}^{(4)}= \frac{1}{3}
\beta^{2}\cosh
(\sqrt{2}\beta \phi_{0}),
\end{equation}
where $\phi_{0}$ is  the background solution to the equation of
motion of the model. Similarly, the boundary three and four point
couplings may be found as:
\begin{equation}
C_{boundary}^{(3)}=\frac{\sqrt{2}
\beta}{12}\left(
\sigma_{1}e^{\beta \phi_{0}/\sqrt{2}}-\sigma_{0}e^{-\beta
\phi_{0}/\sqrt{2}}\right)
\end{equation}
\begin{equation}
C_{boundary}^{(4)}=\frac{\beta^{2}}{48}\left(\sigma_{1}
e^{\beta
\phi_{0}/\sqrt{2}}+\sigma_{0} e^{-\beta\phi_{0}/\sqrt{2}}
\right).
\end{equation}
Meanwhile, we may consider the linear perturbation of the equation
of the motion and the boundary condition around the background
field \cite{CDR,C} to yield
\begin{equation}\label{background}
e^{\beta \phi_{0}/\sqrt{2}}=\frac{1+e^{2(x-x_{0})}}{1-e^{2(x-x_{0})}},
\end{equation}
where  $x_{0}$ is related to the boundary data and satisfies
\begin{equation}\label{boundary parameter}
\coth x_{0}=\sqrt{\frac{1+\sigma_{0}}{1+\sigma_{1}}}.
\end{equation}
So, after some manipulation, the bulk three- and four-point
couplings become
\begin{equation}
C_{bulk}^{(3)}=\frac{4\sqrt{2}}{3} \beta\cosh 2(x-x_{0})\left(
\coth^{2}2(x-x_{0}) -1\right),
\end{equation}

\begin{equation}
C_{bulk}^{(4)}=\frac{1}{3} \beta^{2}\left( 2 \coth^{2}
2(x-x_{0})-1 \right)
\end{equation}
and similarly,
\begin{equation}
C_{boundary}^{(3)}=\frac{\sqrt{2}
\beta}{12}\left(
\sigma_{1}\coth x_{0}-\sigma_{0}\tanh x_{0}\right),
\end{equation}
\begin{equation}
C_{boundary}^{(4)}=\frac{\beta^{2}}{48}\left(\sigma_{1}
\coth x_{0} +\sigma_{0}\tanh x_{0}\right).
\end{equation}

We also need to find the Green's function for the theory. The
two-point propagator for the boundary sinh-Gordon model is known
\cite{C} to have the form
\begin{eqnarray}
G(x,t;x',t')&=&\int \int \frac{d\omega}{2\pi} \frac{dk}{2\pi}
\frac{ie^{-i \omega (t-t')} }{\omega^{2}- k^{2}-4 +i \rho}
\left(f(k,x)f(-k,x') e^{ik(x-x')}
\right. \nonumber\\
& & \hspace{1.35in}+\left.K_{c}f(-k,x)f(-k,x')e^{-ik(x+x')}\right),
\end{eqnarray}
where
\begin{equation}
f(k,x)=\frac{ik-2 \coth 2(x-x_{0})}{ik+2}
\end{equation}
and $K_{c}$ represent  the classical reflection factor of the
model which is given by
\begin{equation}\label{classical reflection}
K_{c}=\left(\frac{(ik)^{2}+2ik\sqrt{1+\sigma_{0}}\sqrt{1+\sigma_{1}}
+2(\sigma_{0}+\sigma_{1})}{(ik)^{2}-2ik\sqrt{1+\sigma_{0}}
\sqrt{1+\sigma_{1}}+
2(\sigma_{0}+\sigma_{1})}\right)\left(\frac{ik-2}{ik+2}\right).
\end{equation}
Indeed, the quantum reflection factor (2.8) reduces \cite{CDR} to
the classical reflection (3.13) in the classical limit  $\beta
\rightarrow 0$ provided  $E(\sigma_{0},\sigma_{1}, 0)=
a_{0}+a_{1}$ and $F(\sigma_{0},\sigma_{1}, 0)=a_{0}-a_{1}$.

In this paper it is intended to calculate the quantum corrections
to the classical reflection factor of the sinh-Gordon model at one
loop order and up to the second order in the difference of the
boundary parameters. Hence, it is necessary to expand the bulk
and the boundary couplings. In fact,
\begin{equation}
C_{bulk}^{(3)}=\frac{2\sqrt{2}}{3}\beta
\frac{\epsilon}{1+\sigma_{1}}e^{2x}+...
\end{equation}

\begin{equation}
C_{bulk}^{(4)}=\frac{1}{3}\beta^{2}\left(1+\frac{1}{2}\frac{\epsilon^{2}}
{(1+\sigma_{1})^{2}}e^{4x}+...\right),
\end{equation}
where $\epsilon=\sigma_{0}-\sigma_{1}$.
Similarly,
\begin{equation}
C_{boundary}^{(3)}=\frac{\sqrt{2}\beta}{12}\left(-\frac{\epsilon}{1+\sigma_{1}}
\right)+...
\end{equation}
and
\begin{equation}
C_{boundary}^{(4)}=\frac{\beta^{2}}{48}
\left(2\sigma_{1}+\epsilon-\frac{\sigma_{1}+2}{4(1+\sigma_{1})^{2}}
\epsilon^{2} \right)+....
\end{equation}
Now let us find the expansions of  $f(k,x)$ and the classical
reflection factor $K_{c}$, both of them appear in the Green's
function (3.11). These can be shown to be
\begin{equation}
f(k,x)=1+\frac{1}{4}\frac{\epsilon^{2}}{(1+\sigma_{1})^{2}}\frac{1}{ik+2}
e^{4x}+...
\end{equation}
and
\begin{eqnarray}
& &K_{c}=\frac{ik+2\sigma_{1}}{ik-2\sigma_{1}}
+\frac{2ik}{(ik-2\sigma_{1})^{2}}\epsilon \nonumber\\ &
&\,\,\,\,\,+\frac{1}{2}\frac{ik(ik^{3}-4k^{2}-6k^{2}\sigma_{1}-4ik\sigma_{1}+8
\sigma_{1}^{2}-16 -16\sigma_{1})}
{(1+\sigma_{1})(ik-2)(ik+2)(ik-2\sigma_{1})^{3}} \epsilon^{2}
\end{eqnarray}
or in a compact form
\begin{equation}
K=K_{0}+K_{1}\epsilon+K_{2}\epsilon^{2}+....
\end{equation}
Here, $K_{0}$ is the classical reflection factor when the two
boundary parameters are equal. To calculate the one loop ($ O(\beta^{2})$) quantum
corrections to the classical reflection factor, we use the idea
introduced by Kim \cite{K1} and developed by Corrigan \cite{C}. In
other words, after perturbative calculation of the propagator  the
$ O(\beta^{2})$ corrections to the classical reflection factor can
be made by looking at the coefficient of $e^{-ik(x+x')}$ in the
residue of the on-shell pole in the asymptotic region
$x,x'\rightarrow -\infty$.

The computations corresponding to the one loop quantum corrections
to the classical reflection factor may be performed by the
standard perturbation theory which has been generalised to the
affine Toda field theory\cite{C,K1,K2,T} in the presence of an
integrable boundary. There are three basic types of Feynman
diagrams which contribute to the two-point Green's function  of
affine Toda field theory in one loop order. These are shown in
figure 1. Now by looking at the three-point and four-point
couplings, we realize that all types of these diagrams are
concerned in our problem ( up to the second order in the
difference of the boundary data). In what follows we calculate the
contributions of types I and III Feynman diagrams to the
reflection factor. The remaining diagrams will be treated
elsewhere.
\vspace{.5in}
\begin{center}
\includegraphics
{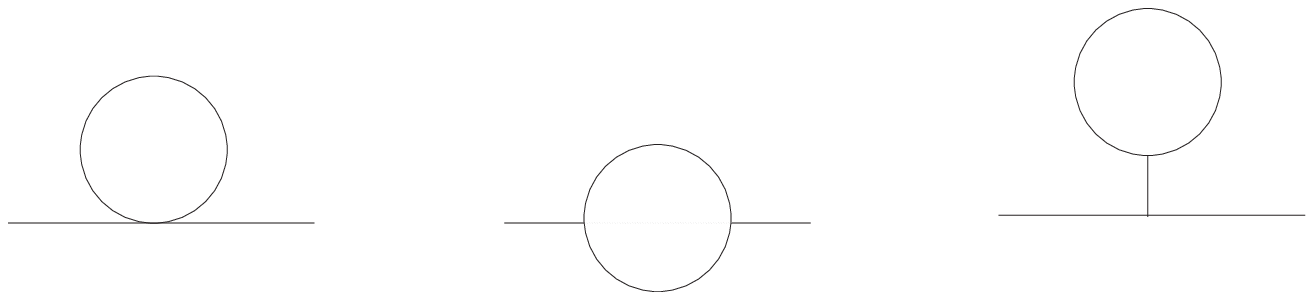}
\vspace{.25in}

\end{center}

\hspace{.73in}I\hspace{1.85in} II\hspace{1.75in} III

\vspace{.25in}
\hspace{.5in}Figure 1: Three basic Feynman diagrams in one loop
order.
\vspace{.25in}

\resection{Type III (boundary-boundary)}

 Clearly,  the type III Feynman diagram includes four different
configurations  depending on the  fact that the interaction
vertices to be located  in the bulk region or at the boundary.
This section deals with the calculations corresponding to the
contribution of  the type III Feynman diagram to the reflection
factor  when both  vertices are placed at the boundary. The
associated contribution may be given by
\begin{equation}
-\frac{\beta^{2}}{4}\frac{\epsilon^{2}}{(1+\sigma_{1})^{2}} \int
\int dt dt'G(x_{1},t_{1};0,t)G(0,t;0,t')
G(0.t';0,t')G(0,t;x_{2},t_{2}).
\end{equation}
Here, the related three point couplings and the combinatorial
factor have been considered. It is evident that in this case the
two-point Green function has the simplest form as
\begin{equation}
G(x,t;x',t')=i\int\int
\frac{d\omega}{2\pi}\frac{dk}{2\pi}\frac{e^{-i\omega(t-t')}}{\omega^{2}-k^{2}-4+i\rho}
\left(e^{ik(x-x')}+K_{0}(k) e^{-ik(x+x')}\right),
\end{equation}
where
\begin{equation}
K_{0}(k)=\frac{ik+2\sigma_{1}}{ik-2\sigma_{1}}.
\end{equation}
Moreover, the loop propagator has been found  in Ref.\cite{C} and
so,
\begin{equation}
G(0,t';0,t')=-a_{1}\frac{\cos a_{1}\pi}{\sin a_{1}\pi}.
\end{equation}
Now it is convenient to calculate the time integral of the other
middle propagator in (4.1), i.e.
\begin{equation}
\int dt' G(0,t ;0,t')
\end{equation}
which is equal to
\begin{equation}
i\int dt'\int\int
\frac{d\omega''}{2\pi}\frac{dk''}{2\pi}\frac{e^{-i\omega''(t-t')}}{\omega''^{2}-k''^{2}-4+i\rho}
\left(1+\frac{ik''+2\sigma_{1}}{ik''-2\sigma_{1}}\right).
\end{equation}
 Integrating over $t'$ gives us a Dirac delta function so, the
 above relation reduces to

\begin{equation}
i\int \frac{dk''}{2\pi}\left(\frac{1}{-k''^{2}-4}\right)
\left(\frac{2ik''}{ik''-2\sigma_{1}}\right).
\end{equation}
Hence,
\begin{equation}
\int dt'G(0,t ;0,t')=-\frac{i}{2(1+\sigma_{1})}.
\end{equation}

Therefore, up to now the contribution (4.1)  has the form
\begin{eqnarray}
& &\frac{i\beta^{2}}{8}\frac{a_{1}\cos a_{1}\pi}{\sin
a_{1}\pi}\frac{\epsilon^{2}}{(1+\cos a_{1}\pi)^{3}} \int dt
\int\int
\frac{d\omega}{2\pi}\frac{dk}{2\pi}\frac{e^{-i\omega(t_{1}-t)}}{\omega^{2}-k^{2}-4+i\rho}
\left(e^{ikx_{1}}+K_{0}(k) e^{-ikx_{1}}\right) \nonumber\\ &
&\hspace{1in} \times\,\int\int
\frac{d\omega'}{2\pi}\frac{dk'}{2\pi}\frac{e^{-i\omega'(t-t_{2})}}{\omega'^{2}-k'^{2}-4+i\rho}
\left(e^{-ik'x_{2}}+K_{0}(k') e^{-ik'x_{2}}\right).
\end{eqnarray}
First of all, it is understood to perform  the transformation  $k
\rightarrow -k$ in the first term of the first propagator.
Secondly, after integration over $t$ which ensures energy
conservation at the interaction vertex, the result will be a Dirac
delta function which immediately gives rise to the substitution
$\omega=\omega'$. Finally,  the momenta of the two propagator can
be integrated out by taking the
 contours to  be closed in the upper half-plane and considering  the pole at
$\hat{k}=k=k'=\sqrt{\omega^{2}-4}$ and ignoring the other pole
i.e. $-2i\sigma_{1}$ (when $\sigma_{1}<0)$ due to the fact that
its residue will vanish in the limit $x_{1},x_{2} \rightarrow
-\infty$. Thus the type III (boundary-boundary) contribution
becomes
\begin{equation}
\frac{i\beta^{2}}{8}\frac{a_{1}\cos a_{1}\pi}{\sin
a_{1}\pi}\frac{\epsilon^{2}}{(1+\cos a_{1}\pi)^{3}} \int
\frac{d\omega}{2\pi}e^{-i\omega(t_{1}-t_{2})}
e^{-i\hat{k}(x_{1}+x_{2})}\frac{1}{(i\hat{k}-2\sigma_{1})^{2}},
\end{equation}
where $\hat k=2\sinh\theta$.
\resection{Type  III  (boundary-bulk) }

Now let us  calculate the contribution of   the type III Feynman
diagram to the reflection factor when the interaction vertex
corresponding to the loop is placed inside the bulk region and the
other vertex is at the boundary.
 As before, the combinatorial factor of the diagram should be
 considered as a coefficient factor. Moreover, in this case we
 must take into account the bulk three-point coupling in the loop
 vertex as well as the boundary three-point coupling in the other
 vertex.  The contribution this time is
\begin{eqnarray}
& &\frac{2\beta^{2}}{(1+\cos a_{1}\pi)^{2}} \epsilon^{2}
\int_{-\infty}^{0}dx'  \int \int dt dt'G(x_{1},t_{1};0,t )G(0,t
;x',t') \nonumber\\ & &\hspace{2.5in}G(x',t';x',t')G(0,t
;x_{2},t_{2}) e^{2x'}.
\end{eqnarray}
 In fact, in the previous section we found the following
 relation which is some part of the contribution (5.1):
\begin{equation}
\int dt
G(x_{1},t_{1};0,t)G(0,t;x_{2},t_{2})=-\int\frac{d\omega}{2\pi}e^{-i\omega(t_{1}-t_{2})}
e^{-i\hat{k}(x_{1}+x_{2})}\frac{1}{(i\hat{k}-2\sigma_{1})^{2}}.
\end{equation}
 Clearly, in the boundary-bulk contribution (5.1), it is seen that the $t'$
 dependence is involved only in the propagator $G(0,t,x',t')$. So,
 it is convenient to compute the following relation
\begin{equation}
\int dt' G(0,t;x',t')= i\int dt'\int\int
\frac{d\omega'}{2\pi}\frac{dk'}{2\pi}
\left(\frac{e^{-i\omega'(t-t')}}{\omega'^{2}-k'^{2}-4+i\rho}\right)
\left(\frac{2ik'}{ik'-2\sigma_{1}}\right)e^{-ik'x'}.
\end{equation}
Integrating over $t'$ generates a Dirac delta function which
allows the substitution $\omega' = 0$ and hence,
\begin{equation}
\int dt' G(0,t;x',t')=
i\int\frac{dk'}{2\pi}\left(\frac{e^{-ik'x'}}{-k'^{2}-4}\right)
\left(\frac{2ik'}{ik'-2\sigma_{1}}\right).
\end{equation}
Therefore,  we obtain
\begin{equation}
\int dt' G(0,t;x',t')=-\frac{i}{2(1+\sigma_{1})}e^{2x'}.
\end{equation}

Now let us  calculate the loop propagator which has the form
\begin{equation}
G(x',t';x',t')=i\int\int
\frac{d\omega''}{2\pi}\frac{dk''}{2\pi}\frac{1}{\omega''^{2}-k''^{2}-4+i\rho}
\left(1+K_{0}(k'')e^{-2ik''x'}\right).
\end{equation}
In fact, the above integral is divergent. However, this divergence
can be removed by an infinite renormalization of the mass
parameter in the bulk potential. In other words, a minimal
subtraction of the divergent part can be made by adding an
appropriate counterterm to the bulk potential. Meanwhile, if we
integrate over $\omega''$ then, we will find

\begin{equation}
G(x',t';x',t')=\frac{1}{2}\int
\frac{dk''}{2\pi}\frac{1}{\sqrt{k''^{2}+4}}
\frac{ik''+2\sigma_{1}}{ik''-2\sigma_{1}}e^{-2ik''x'}.
\end{equation}
Hence, the remaining part of the contribution  (5.1)  is
simplified as (apart from the coefficients)
\begin{eqnarray}
& &\int \int dt'dx'G(0,t;x',t')G(x',t';x',t')e^{2x'} \nonumber\\ &
&\,\,\,=-\frac{i}{4(1+\sigma_{1})} \int_{-\infty}^{0} dx'\int
\frac{dk''}{2\pi}\frac{1}{\sqrt{k''^{2}+4}}
\frac{ik''+2\sigma_{1}}{ik''-2\sigma_{1}}e^{(4-2ik'')x'}
\end{eqnarray}
In order to integrate over $x'$, one may use the following formula
\begin{equation}
\int_{-\infty}^{0} dx' e^{i k'' x' +\tau x'}=\frac{-i}{k''-i \tau}
\end{equation}
where the quantity $\tau$ must be positive.
 The $k''$ integration can be evaluated by closing the contour in
 the upper (lower) half-plane depending on whether $\sigma_{1} >
 0$ ($\sigma_{1} <0$). However, due to the branch cut the contour has to run
 around the cut line. Note,  the cut line stretches from $k''=2i$ to infinity along
 the imaginary axis when $\sigma_{1} >
 0$. Otherwise, the cut line run from $k''=-2i$ to $-\infty$
 along the imaginary axis when $\sigma_{1} <
 0$. So,  after performing the required integrations
\begin{eqnarray}
& &\int \int dt'dx'G(0,t;x',t')G(x',t';x',t')e^{2x'} \nonumber\\ &
& \, \, \,\,=\frac{i}{16\pi(1+\cos a_{1}\pi)} \left( \frac{2\pi
a_{1}\cos a_{1}\pi}{(1-\cos a_{1}\pi)\sin a_{1}\pi} -\frac{1+\cos
a_{1}\pi}{1-\cos a_{1}\pi}\right).
\end{eqnarray}
Therefore, the contribution of the type III (boundary-bulk)
diagram to the reflection factor is
\begin{eqnarray}
& &-\frac{i\beta^{2}\epsilon^{2}}{8\pi(1+\cos a_{1}\pi)^{3}}
\left( \frac{2\pi a_{1}\cos a_{1}\pi}{(1-\cos a_{1}\pi)\sin
a_{1}\pi} -\frac{1+\cos a_{1}\pi}{1-\cos a_{1}\pi}\right)
\nonumber\\ & &\int \frac{d\omega}{2\pi}e^{-i\omega(t_{1}-t_{2})}
e^{-i\hat{k}(x_{1}+x_{2})}\frac{1}{(i\hat{k}-2\sigma_{1})^{2}}.
\end{eqnarray}

\resection{Type   III  (bulk-boundary) }

Let us consider the type III Feynman  diagram when the loop vertex
is at the boundary and the other interaction vertex  is  inside
the bulk region. So, in this case the contribution  is
\begin{eqnarray}
& &\frac{2\beta^{2}}{(1+\cos a_{1}\pi)^{2}} \epsilon^{2}
\int_{-\infty}^{0}dx  \int \int dt
dt'G(x_{1},t_{1};x,t)G(x,t;0,t') \nonumber\\ &
&\hspace{2.4in}\times\,G(0,t';0,t')G(x,t;x_{2},t_{2}) e^{2x}.
\end{eqnarray}

In fact, we have obtained the two middle propagators  in the
previous sections. Hence,

\begin{equation}
G(0,t';0,t')=-a_{1}\frac{\cos a_{1}\pi}{\sin a_{1}\pi}
\end{equation}
and
\begin{equation}
\int dt' G(x,t;0,t')=-\frac{i}{2(1+\sigma_{1})}e^{2x}.
\end{equation}
So, the contribution (6.1)  can be shown in detail as (after
doing the integration over $t$)

\begin{eqnarray}
& &\frac{i\beta^{2}\epsilon^{2}}{(1+\cos a_{1}\pi)^{3}}
 \frac{a_{1}cos a_{1}\pi}{\sin
a_{1}\pi} \nonumber\\ & &\times\,\int_{-\infty}^{0}dx \int\int
\frac{d\omega}{2\pi}\frac{dk}{2\pi}\frac{ie^{-i\omega(t_{1}-t_{2})}}{\omega^{2}-k^{2}-4+i\rho}
\left(e^{-ik(x_{1}-x)}+K_{0}(k) e^{-ik(x_{1}+x)}\right)
\nonumber\\ & &\hspace{.4in} \times \,\int
\frac{dk'}{2\pi}\frac{i}{\omega^{2}-k'^{2}-4+i\rho}
\left(e^{ik'(x-x_{2})}+K_{0}(k') e^{-ik'(x+x_{2})}\right)e^{4x}.
\end{eqnarray}
The integration over $x$ may be done by using the device (5.9).
Meanwhile, it is necessary to perform a transformation  $k
\rightarrow -k$ in the first term of the first propagator.
Therefore, the above expression becomes
\begin{eqnarray}
& &\frac{i\beta^{2}\epsilon^{2}}{(1+\cos a_{1}\pi)^{3}}
 \frac{a_{1}\cos a_{1}\pi}{\sin
a_{1}\pi} \nonumber\\ & &\int \int\int
\frac{d\omega}{2\pi}\frac{dk}{2\pi} \frac{dk'}{2\pi}
e^{-i\omega(t_{1}-t_{2})}e^{-i(kx_{1}+k'x_{2})}
\frac{i}{\omega^{2}-k^{2}-4+i\rho}
\,\frac{i}{\omega^{2}-k'^{2}-4+i\rho} \nonumber\\ & &\,\, \times
\,
\left\{-\frac{i}{k+k'-4i}-\frac{iK_{0}(k')}{k-k'-4i}-\frac{iK_{0}(k)}{-k+k'-4i}
-\frac{iK_{0}(k)K_{0}(k')}{-k-k'-4i}\right\}.
\end{eqnarray}
Clearly, all that remains  is to integrate over the momenta $k$,
$k'$. This can be achieved   by completing the contours in the
upper half-plane and picking up the pole at
$\hat{k}=k=k'=\sqrt{\omega^{2}-4}$ and regarding the  fact that
all the other poles have no contributions because their residues
yield exponentially damped terms in the asymptotic region
$x_{1},x_{2} \rightarrow -\infty $. Hence, the contribution of the
type III (bulk-boundary) Feynman diagram to the reflection factor
is
\begin{eqnarray}
& &\frac{\beta^{2}\epsilon^{2}}{(1+\cos a_{1}\pi)^{3}}
 \frac{a_{1}\cos a_{1}\pi}{\sin
a_{1}\pi} \int \frac{d\omega}{2\pi}
 e^{-i\omega(t_{1}-t_{2})}
e^{-i\hat{k}(x_{1}+x_{2})}\frac{1}{(2\hat{k})^{2}}
 \nonumber\\
& & \hspace{1in} \times
\,\left\{\frac{1}{2\hat{k}-4i}+\frac{i}{2}K_{0}(\hat{k})
-\frac{1}{2\hat{k}+4i}K_{0}^{2}(\hat{k}) \right\}.
\end{eqnarray}

\resection{Type   III  (bulk-bulk) }

In this section we study the
  type III (bulk-bulk) Feynman diagram which contribute to the
  quantum correction to the classical reflection factor. It is
  evident that in this case both interaction vertices are located
  inside the bulk region. So, in this configuration the
  corresponding contribution is given by
\begin{eqnarray}
& &-16\frac{\beta^{2}}{(1+\cos a_{1}\pi)^{2}} \epsilon^{2}
\int_{-\infty}^{0}dx \int_{-\infty}^{0}dx'  \int \int dt
dt'G(x_{1},t_{1};x,t)G(x,t;x',t') \nonumber\\ &
&\hspace{2.5in}G(x',t';x',t')G(x,t;x_{2},t_{2}) e^{2x}e^{2x'}.
\end{eqnarray}
As  it was shown in the type III (boundary-bulk) case, the loop
propagator can be simplified to obtain

\begin{equation}
G(x',t';x',t')=\frac{1}{2}\int
\frac{dk_{1}}{2\pi}\frac{1}{\sqrt{k_{1}^{2}+4}}
\left(\frac{ik_{1}+2\sigma_{1}}{ik_{1}-2\sigma_{1}}\right)
e^{-2ik_{1}x'}.
\end{equation}
Moreover, the time variable $t'$ appears only in the other middle
propagator.  Therefore,  it is appropriate to calculate the
following relation

\begin{eqnarray}
\int dt'G(x,t;x',t')=i\int\int\int dt'
\frac{d\omega''}{2\pi}\frac{dk''}{2\pi}\frac{e^{-i\omega''(t-t')}}{\omega''^{2}-k''^{2}
-4+i\rho} \left(e^{ik''(x-x')}\right.\nonumber\\
\left.\hspace{2.5in}+K_{0}(k'') e^{-ik''(x+x')}\right).
\end{eqnarray}
The integration over $t'$ gives us a Dirac delta function which
replaces $\omega''$ by zero and hence,
\begin{equation}
\int dt'G(x,t;x',t')=\int
\frac{dk''}{2\pi}\left(\frac{i}{-k''^{2}-4}\right) e^{-ik''x'}
\left(e^{ik''x}+K_{0}(k'') e^{-ik''x}\right).
\end{equation}

Let us rewrite the contribution (7.1) in the expanded form  in
 order to find how we may carry out the required integrations (
first setting $k \rightarrow -k$ in the first term of the first
propagator)

\begin{eqnarray}
& &16\frac{\beta^{2}}{(1+\cos a_{1}\pi)^{2}} \epsilon^{2}
\int_{-\infty}^{0}dx \int_{-\infty}^{0}dx' \int dt \int\int
\frac{d\omega}{2\pi}\frac{dk}{2\pi}\frac{ie^{-i\omega(t_{1}-t)}}{\omega^{2}-k^{2}-4+i\rho}
\left(e^{-ik(x_{1}-x)} \right.\nonumber\\ & &\left.
\hspace{3in}+K_{0}(k) e^{-ik(x_{1}+x)}\right) \nonumber\\ &
&\times \,\int \frac{dk''}{2\pi}\frac{i}{-k''^{2}-4} e^{-ik''x'}
\left(e^{ik''x}+K_{0}(k'') e^{-ik''x}\right)e^{2x}e^{2x'}
\nonumber\\ & &\times \,\frac{1}{2}\int
\frac{dk_{1}}{2\pi}\frac{1}{\sqrt{k_{1}^{2}+4}}
\frac{ik_{1}+2\sigma_{1}}{ik_{1}-2\sigma_{1}}e^{-2ik_{1}x'}
\nonumber\\ & &\times \,\int\int
\frac{d\omega'}{2\pi}\frac{dk'}{2\pi}\frac{ie^{-i\omega(t-t_{2})}}{\omega'^{2}-k'^{2}-4+i\rho}
\left(e^{ik'(x-x_{2})}+K_{0}(k') e^{-ik'(x+x_{2})}\right).
\end{eqnarray}
The starting point is to perform the $t$ integration which
immediately removes the energy variable $\omega'$. Moreover, by
multiplying the first and the third propagator then we obtain four
pole pieces. If we perform the calculations corresponding to one
of them ( for example the first one which involves $e^{i(k+k')x}$
term),  then  the calculations for the  remaining three pole pieces
may be done  in the same manner, except that $k+k'$ is replaced by
one of $k-k'$, $-k+k'$ and  $-k-k'$. So, in what follows we
continue the computations in detail only   for one pole piece.
Now,
 the integration over $x'$ is achieved by means of the device (5.9) and gives
\begin{equation}
\int_{-\infty}^{0} dx' e^{\left(2-i(k''+2k_{1})\right)x'}=
\frac{i}{k''+2k_{1}+2i}.
\end{equation}
Similarly, the  $x$ integration can be  done and yields
\begin{eqnarray}
& &\int_{-\infty}^{0} dx e^{\left(2+i(k+k')\right)x}
\left(e^{ik''x}+K_{0}(k'')e^{-ik''x}\right) \nonumber\\ & & \, \,
\,=-\frac{i}{k+k'+k''-2i}-\frac{i}{k+k'-k''-2i}K_{0}(k'').
\end{eqnarray}

So, we find that we must solve the following integral
\begin{eqnarray}
& &16\frac{\beta^{2}\epsilon^{2}}{(1+\cos a_{1}\pi)^{2}}
\int\int\int \frac{d\omega}{2\pi}\frac{dk}{2\pi}\frac{dk'}{2\pi}
\frac{e^{-i\omega(t_{1}-t_{2})}}{\omega^{2}-k^{2}-4+i\rho}
\,\frac{e^{-i(kx_{1}+k'x_{2})}}{\omega^{2}-k'^{2}-4+i\rho}
\nonumber\\ & &\times \,\frac{1}{2}\int \int\frac{dk''}{2\pi}
\frac{dk_{1}}{2\pi}\frac{1}{\sqrt{k_{1}^{2}+4}}
\left(\frac{ik_{1}+2\sigma_{1}}{ik_{1}-2\sigma_{1}}\right)
\left(\frac{i}{k''^{2}+4}\right)\left(\frac{1}{-k''-2k_{1}-2i}\right)
\nonumber\\ & &\times
\,\left(\frac{1}{k+k'+k''-2i}+\frac{1}{k+k'-k''-2i}K_{0}(k'')\right).
\end{eqnarray}
 It turns out to be more convenient to integrate over $k''$ then
 afterwards over
$k_{1}$. Hence, first of all, we need to evaluate
\begin{equation}
\int\frac{dk''}{2\pi}\frac{i}{k''^{2}+4}\frac{1}{-k''-2k_{1}-2i}
\left(\frac{1}{k+k'+k''-2i}+\frac{1}{k+k'-k''-2i}K_{0}(k'')\right).
\end{equation}
The above integral may be solved by  closing the contours in the
upper or lower half-plane in order to become free  of all extra
poles except  $\pm 2i$. Therefore,  after some manipulation we
obtain
\begin{eqnarray}
& &
\int\frac{dk''}{2\pi}\frac{i}{k''^{2}+4}\frac{1}{-k''-2k_{1}-2i}
\left(\frac{1}{k+k'+k''-2i}+\frac{1}{k+k'-k''-2i}K_{0}(k'')\right)
\nonumber\\ & &\, \, \,=
\frac{1}{4i}\left\{\frac{1-\sigma_{1}}{1+\sigma_{1}}\,\frac{1}{2k_{1}+4i}
\,\frac{1}{k+k'-4i} +\frac{1}{2k_{1}+4i}\,\frac{1}{k+k'-2k_{1}-4i}
\right.\nonumber\\ & & \left.\hspace{2.4in}
-\frac{1}{k+k'-4i}\,\frac{1}{k+k'-2k_{1}-4i}\right\}.
\end{eqnarray}

Next, we need to calculate   the  integration over $k_{1}$. In
order to calculate  the loop momentum  $k_{1}$ integral, we will have
different kinds of integrals however, all of them can be solved
similarly. For example, consider the integral
\begin{equation}
\int \frac{dk_{1}}{\sqrt{k_{1}^{2}+4}}\frac{1}{k+k'-2k_{1}-4i}.
\end{equation}
Let us close the contour in the upper half-plane and due to the
branch cut which  extends  from $2i$ to infinity along the
imaginary axis, the contour has to turn around the cut line.
Therefore,   the  integral (7.11)  reduces to
\begin{equation}
2\int_{2}^{\infty}\frac{dy}{\sqrt{y^{2}-4}}\frac{1}{k+k'-2iy-4i}
\end{equation}
and the above integral gives the following result
\begin{eqnarray}
& &\int \frac{dk_{1}}{\sqrt{k_{1}^{2}+4}}\frac{1}{k+k'-2k_{1}-4i}
\nonumber\\ & &=\frac{1}{\sqrt{\frac{(k+k'-4i)^{2}}{4}+4}}
\ln{\left\{\frac{1+\frac{i}{4}(k+k'-4i)+\frac{i}{2}\sqrt{\frac{(k+k'-4i)^{2}}{4}+4}}
{1+\frac{i}{4}(k+k'-4i)-\frac{i}{2}\sqrt{\frac{(k+k'-4i)^{2}}{4}+4}}\right\}}.
\end{eqnarray}

Now we can  write down the solution of the $k_{1}$ integration as
\begin{eqnarray}
& & \frac{1}{8i}\int
\frac{dk_{1}}{2\pi}\frac{1}{\sqrt{k_{1}^{2}+4}}
\left(\frac{ik_{1}+2\sigma_{1}}{ik_{1}-2\sigma_{1}}\right)
\left\{\frac{1-\sigma_{1}}{1+\sigma_{1}}\,\frac{1}{2k_{1}+4i}\frac{1}{k+k'-4i}
\right.\nonumber\\ & &\hspace{1in}
\left.+\frac{1}{2k_{1}+4i}\frac{1}{k+k'-2k_{1}-4i}
 -\frac{1}{k+k'-4i}\frac{1}{k+k'-2k_{1}-4i}\right\}
\nonumber\\ & &  \, \, =-\frac{k+k'-2i-2i\cos a_{1}\pi}{16
\pi(1-\cos a_{1}\pi)(k+k')(k+k'-4i)} \nonumber\\ & &  \, \, \,
+\frac{a_{1}\cos a_{1}\pi}{16\sin^{3} a_{1}\pi}
\,\frac{2k+2k'+4i\cos a_{1}\pi -12i} {(k+k'-4i)(k+k'+4i\cos
a_{1}\pi-4i)} \nonumber\\ & & \, \, \,
-\frac{1}{4\pi(k+k')(k+k'-4i)}\,\frac{k+k'-4i\cos a_{1}\pi
-4i}{k+k'+4i\cos a_{1}\pi-4i}
\,\frac{1}{\sqrt{\frac{(k+k'-4i)^{2}}{4}+4}} \nonumber\\ & &
\hspace{1.5in} \times \,
\ln{\left\{\frac{1+\frac{i}{4}(k+k'-4i)+\frac{i}{2}\sqrt{\frac{(k+k'-4i)^{2}}{4}+4}}
{1+\frac{i}{4}(k+k'-4i)-\frac{i}{2}\sqrt{\frac{(k+k'-4i)^{2}}{4}+4}}\right\}}.
\end{eqnarray}
Finally, regarding the type III (bulk-bulk) contribution (7.5),
all that remains is to integrate over the momenta  $k,k'$. This
can be achieved by means of the contours in the upper half-plane
and considering the pole at $\hat{k}=k=k'=\sqrt{\omega^{2}-4}$ and
ignoring all the other extra poles because their contributions
will vanish when $x_{1},x_{2}\rightarrow -\infty$. Hence, the
relation (7.5)
  has the  solution

\begin{eqnarray}
& &-\frac{\beta^{2}}{(1+\cos a_{1}\pi)^{2}} \epsilon^{2} \int
\frac{d\omega}{2\pi}
e^{-i\omega(t_{1}-t_{2})}e^{-i\hat{k}(x_{1}+x_{2})}\frac{1}{(2\hat{k})^{2}}
\nonumber\\ & & \left\{ \left(-\frac{\hat{k}-i-i\cos
a_{1}\pi}{\pi(1- \cos a_{1}\pi)\hat{k}(2\hat{k}-4i)}
+\frac{a_{1}\cos a_{1}\pi}{\sin^{3} a_{1}\pi}
\,\frac{\hat{k}+i\cos a_{1}\pi -3i} {(\hat{k}-2i)(\hat{k}+2i\cos
a_{1}\pi-2i)} \right. \right. \nonumber\\ & &\hspace{1in} \left.
\left. -\frac{1}{\pi\hat{k}(\hat{k}-2i)}\,\frac{\hat{k}-2i\cos
a_{1}\pi -2i}{\hat{k}+2i \cos a_{1}\pi-2i}
\,\frac{1}{\sqrt{(\hat{k}-2i)^{2}+4}} \right.\right.\nonumber\\ &
&\hspace{2in}\left.\left. \times
\,\ln{\left\{\frac{1+\frac{i}{2}(\hat{k}-2i)+\frac{i}{2}\sqrt{(\hat{k}-2i)^{2}+4}}
{1+\frac{i}{2}(\hat{k}-2i)-\frac{i}{2}\sqrt{(\hat{k}-2i)^{2}+4}}\right\}}
\right) \right.\nonumber\\ & &\left. + K_{0}(\hat{k})
\left(-\frac{i}{2\pi(1-\cos a_{1}\pi)^{2}} -\frac{ia_{1}\cos
a_{1}\pi(\cos a_{1}\pi-3)} {2\sin^{3} a_{1}\pi(1-\cos a_{1}\pi)}
+\frac{i(1+\cos a_{1}\pi)}{12\pi (1-\cos a_{1}\pi)}\right)
\right.\nonumber\\ & & \left. + K_{0}^{2}(\hat{k})
\left(\frac{\hat{k}+i+i\cos a_{1}\pi}{\pi(1- \cos
a_{1}\pi)\hat{k}(2\hat{k}+4i)} -\frac{a_{1}\cos a_{1}\pi}{\sin^{3}
a_{1}\pi} \,\frac{\hat{k}-i\cos a_{1}\pi +3i}
{(\hat{k}+2i)(\hat{k}-2i\cos a_{1}\pi+2i)} \right. \right.
\nonumber\\ & & \hspace{1in}\left. \left.
-\frac{1}{\pi\hat{k}(\hat{k}+2i)}\,\frac{\hat{k}+2i\cos a_{1}\pi
+2i}{\hat{k}-2i \cos a_{1}\pi+2i}
\,\frac{1}{\sqrt{(\hat{k}+2i)^{2}+4}} \right.\right.\nonumber\\ &
&\hspace{1.7in}\left.\left.\times
\,\ln{\left\{\frac{1-\frac{i}{2}(\hat{k}+2i)+\frac{i}{2}\sqrt{(\hat{k}
+2i)^{2}+4}}
{1-\frac{i}{2}(\hat{k}+2i)-\frac{i}{2}\sqrt{(\hat{k}+2i)^{2}+4}}\right\}}
\right) \right\}.
\end{eqnarray}

Notice that the above relation depends on the  $ln$ terms in a
manner which is very inconvenient for a comparison with Ghoshal's
formula. Fortunately, these terms will be cancelled by matching
terms in the contribution corresponding to the  type I (bulk)
Feynman diagram. This is  one of the interesting results which may
be found  in this paper. So, the type III (bulk-bulk) contribution
takes the simple form

\begin{eqnarray}
& &-\frac{\beta^{2}}{(1+\cos a_{1}\pi)^{2}} \epsilon^{2} \int
\frac{d\omega}{2\pi}
e^{-i\omega(t_{1}-t_{2})}e^{-i\hat{k}(x_{1}+x_{2})}\frac{1}{(2\hat{k})^{2}}
\nonumber\\ & & \left\{ \left(-\frac{\hat{k}-i-i\cos
a_{1}\pi}{\pi(1- \cos a_{1}\pi)\hat{k}(2\hat{k}-4i)}
+\frac{a_{1}\cos a_{1}\pi}{\sin^{3} a_{1}\pi}
\,\frac{\hat{k}+i\cos a_{1}\pi -3i} {(\hat{k}-2i)(\hat{k}+2i\cos
a_{1}\pi-2i)} \right) \right. \nonumber\\ & &\left. +
K_{0}(\hat{k})  \left(-\frac{i}{2\pi(1-\cos a_{1}\pi)^{2}}
-\frac{ia_{1}\cos a_{1}\pi(\cos a_{1}\pi-3)} {2\sin^{3}
a_{1}\pi(1-\cos a_{1}\pi)} +\frac{i(1+\cos a_{1}\pi)}{12\pi
(1-\cos a_{1}\pi)}\right) \right.\nonumber\\ & & \left. +
K_{0}^{2}(\hat{k})  \left(\frac{\hat{k}+i+i\cos a_{1}\pi}{\pi(1-
\cos a_{1}\pi)\hat{k}(2\hat{k}+4i)} -\frac{a_{1}\cos
a_{1}\pi}{\sin^{3} a_{1}\pi} \,\frac{\hat{k}-i\cos a_{1}\pi +3i}
{(\hat{k}+2i)(\hat{k}-2i\cos a_{1}\pi+2i)} \right) \right\}.
\nonumber\\
\end{eqnarray}

\resection{Type  I (boundary) }

In fact in the case of the type I Feynman diagram,  there are two
possible configurations  depending on the fact that whether  the
 interaction vertex is  located at the boundary or inside the bulk
region.
 From now on we call these two configurations the  type I (boundary) and
 the type I (bulk) respectively.
  Considering the boundary four point coupling (3.17) and the
associated combinatorial factor as well, the contribution
corresponding to the type I (boundary) diagram is described by
\begin{equation}
-\frac{i\beta^{2}}{4}\left(2\sigma_{1}+\epsilon-\frac{\sigma_{1}+2}
{4(1+\sigma_{1})^{2}}\epsilon^{2}\right) \int_{-\infty}^{\infty}
dt''G(x,t;0,t'')G(0,t'';0,t'')G(0,t'';x',t').
\end{equation}

Let us  find the appropriate form of the two-point Green function
which  will be used many times throughout this and  next sections.
Now by looking at the general form of the propagator (3.11) and
considering the expansions of the classical reflection factor
(3.20) and the function $f(k,x)$ i.e.  (3.18) as well,  the
required form of the two-point Green function will be

\begin{eqnarray}
& &G(x,t;x',t')=i\int\int
\frac{d\omega}{2\pi}\frac{dk}{2\pi}\frac{e^{-i\omega(t-t')}}{\omega^{2}-k^{2}-4+i\rho}
\Biggl\{\left(1+\frac{\epsilon^{2}}{4(1+\sigma_{1})^{2}}\,\frac{1}{(ik+2)}e^{4x}
\right.\Biggr.\nonumber\\ &
&\hspace{2.5in}\left.\Biggl.-\frac{\epsilon^{2}}{4(1+\sigma_{1})^{2}}\,\frac{1}{(
ik-2)}e^{4x'}\right)e^{ik(x-x')} \Biggr.\nonumber\\ &
&\hspace{.6in}\Biggl.
+\biggl(K_{0}-\frac{\epsilon^{2}}{4(1+\sigma_{1})^{2}}\,\frac{1}{(ik-2)}e^{4x}
K_{0}-
\frac{\epsilon^{2}}{4(1+\sigma_{1})^{2}}\,\frac{1}{(ik-2)}e^{4x'}
 K_{0}
\biggr.\Biggr.\nonumber\\ & &\hspace{3in}\biggl.\biggl.+\epsilon
K_{1}+\epsilon^{2} K_{2}\biggr) e^{-ik(x+x')}\Biggr\}.
\end{eqnarray}
So,  in our problem the loop  propagator takes  the following form
\begin{eqnarray}
& &G(0,t'';0,t'')=\int\int
\frac{d\omega''}{2\pi}\frac{dk''}{2\pi}\frac{i}{\omega''-k''^{2}-4+i\rho}
\Bigl(1+K_{0}+\epsilon K_{1}+\epsilon^{2}K_{2} \Bigr.\nonumber\\
& &\Bigl.+\frac{\epsilon^{2}}{4(1+\sigma_{1})^{2}}\frac{1}{(ik''+2)}
-\frac{\epsilon^{2}}{4(1+\sigma_{1})^{2}}\frac{1}{(ik''
-2)}-\frac{\epsilon^{2}}{2(1+\sigma_{1})^{2}}\frac{1}{(ik''-2)}K_{0}
\Bigr).
\end{eqnarray}
First of all it is necessary to perform  a minimal subtraction in
order to remove the divergence of the above integral.  Secondly,
we have solved the integral in Ref. \cite{CC}  up to the first
order in $\epsilon$ . Hence, what we need to do is to solve the
part which is proportional to $\epsilon^{2}$ including  four
terms. Indeed,   three of them can be simply manipulated and
because of this reason we solve  the first  one in detail which is
given by
\begin{equation}
i\int
\int\frac{d\omega''}{2\pi}\frac{dk''}{2\pi}\frac{1}{\omega''-k''^{2}-4+i\rho}
\epsilon^{2} K_{2}(k'')
\end{equation}
or
\begin{eqnarray}
& &\frac{1}{2}\int
\frac{dk''}{2\pi}\frac{1}{\sqrt{k''^{2}+4}}\epsilon^{2}
\left(\frac{i}{4(\sigma_{1}^{2}-2\sigma_{1}+1)(k''+2i)}+\frac{i}
{4(1+\sigma_{1})^{2}(k''-2i)} \right.\nonumber\\ &
&\hspace{.5in}\left.+\frac{i\sigma_{1}(\sigma_{1}^{2}-2\sigma_{1}-1)}
{(1+\sigma_{1})(\sigma_{1}^{3}-\sigma_{1}^{2}-\sigma_{1}+1)}
\,\frac{1}{(k''+2i\sigma_{1})}-\frac{\sigma_{1}^{2}-2}{(1+\sigma_{1})(\sigma_{1}-1)}
\,\frac{1}{(k''+2i\sigma_{1})^{2}} \right.\nonumber\\ &
&\hspace{3in}\left.+\frac{4i\sigma_{1}}{(k''+2i\sigma_{1})^{3}}\right).
\end{eqnarray}

Fortunately, we have already calculated  these kinds of integrals
in Ref. \cite{CC}  except the last one which is (apart from the
coefficients)
\begin{equation}
\int \frac{dk''}{\sqrt{k''^{2}+4}}
\frac{1}{(ik''-2\sigma_{1})^{3}}.
\end{equation}
The above integral can be converted to a complex one and regarding
the branch cut, it reduces to
\begin{equation}
-2\int_{2}^{\infty}\frac{dy}{\sqrt{y^{2}-4}}\frac{1}{(y+2\sigma_{1})^{3}}
\end{equation}
and after   some manipulation we obtain
\begin{equation}
\int \frac{dk''}{\sqrt{k''^{2}+4}}
\frac{1}{(ik''-2\sigma_{1})^{3}} =\frac{3\cos a_{1}\pi}{8\sin
^{4}a_{1}\pi}-\frac{1+2\cos^{2}a_{1}\pi} {8\sin^{5}
a_{1}\pi}a_{1}\pi.
\end{equation}

Now using the above formula and the required formulae in Ref.
\cite{CC},  we may find  the following result
\begin{eqnarray}
& & i
\int\int\frac{d\omega''}{2\pi}\frac{dk''}{2\pi}\frac{1}{\omega''-k''^{2}-4}
\epsilon^{2} K_{2}(k'') \nonumber\\ & &
\hspace{.5in}=\frac{\epsilon^{2}}{4\pi}\left(\frac{1+\cos
a_{1}\pi+\cos^{2}a_{1}\pi}{\sin^{4}a_{1}\pi} -\frac{\cos
a_{1}\pi(2+\cos a_{1}\pi)}{\sin^{5}a_{1}\pi}a_{1}\pi\right).
\end{eqnarray}
Finally after doing the necessary computations and simplifications
in connection with (8.3), we obtain  the loop  propagator as
\begin{eqnarray}
G(0,t'';0,t'')&=&-a_{1}\frac{\cos a_{1}\pi}{\sin a_{1}\pi}
+\epsilon \left(\frac{\cos a_{1}\pi}{2\pi\sin^{2} a_{1}\pi}
-\frac{a_{1}}{2\sin^{3} a_{1}\pi}\right) \nonumber\\ & &
+\epsilon^{2}\left(\frac{2+\cos^{2}
a_{1}\pi}{4\pi\sin^{4}a_{1}\pi} -\frac{3\cos
a_{1}\pi}{4\pi\sin^{5}a_{1}\pi}a_{1}\pi\right).
\end{eqnarray}

Let us consider  the boundary contribution (8.1).  The integration
over  $t''$ creates a Dirac delta function which means  we must
set  $\omega'=\omega$. As before in previous sections, the
remaining integral over the momenta  $k$ and  $k'$ can be achieved
by closing the contours in the upper half-plane and taking into
account the pole at  $k=k'=\hat{k}=\sqrt{\omega^{2}-4}$. Note, all
the other poles will give damped contributions in the asymptotic
region $x,x' \rightarrow -\infty$. So, the type I (boundary)
contribution is
\begin{eqnarray}
&
&-\frac{i\beta^{2}}{4}\left(2\sigma_{1}+\epsilon-\frac{\sigma_{1}+2}
{4(1+\sigma_{1})^{2}}\epsilon^{2}\right) \nonumber\\ &
&\times\, \Biggl\{-\frac{a_{1}\cos a_{1}\pi}{\sin a_{1}\pi} +\epsilon
\left(\frac{\cos a_{1}\pi}{2\pi\sin^{2} a_{1}\pi}
-\frac{a_{1}}{2\sin^{3} a_{1}\pi}\right)
\Biggr. \nonumber\\ & & \hspace{1.1in}\Biggl.
+\epsilon^{2}\left(\frac{2+\cos^{2}
a_{1}\pi}{4\pi\sin^{4}a_{1}\pi} -\frac{3\cos
a_{1}\pi}{4\pi\sin^{5}a_{1}\pi}a_{1}\pi\right)\Biggr\}
 \nonumber\\ & &
\times\,\int \frac{d\omega}{2\pi}e^{-i\omega(t-t')}
e^{-i\hat{k}(x+x')}\frac{1}{(2\hat{k})^{2}} \nonumber\\ &
&\times\,\left\{\frac{2i\hat{k}}{i\hat{k}-2\sigma_{1}}
+\frac{2i\hat{k}}{(i\hat{k}-2\sigma_{1})^{2}}\epsilon
\right.\nonumber\\ & &\hspace{1in}
\left.+\frac{i\hat{k}(-i\hat{k}^{3}+2\hat{k}^{2}-4i\hat{k}\sigma_{1}
+6\hat{k}^{2}\sigma_{1}+16+16\sigma_{1})}
{2(1+\sigma_{1})(i\hat{k}-2)(i\hat{k}+2)(-i\hat{k}+2\sigma_{1})^{3}}
\epsilon^{2}\right\}.
\end{eqnarray}

\resection{Type  I  (bulk) }

Now let us evaluate  the contribution of  the type I (bulk)
Feynman diagram to the classical reflection factor when  the
interaction vertex is placed inside the bulk region. It is clear
that in this case   we have to consider  the bulk four point
coupling (3.15) in the interaction vertex however, the related
combinatorial factor is the same as the boundary case. The
corresponding contribution may be written  as
\begin{eqnarray}
-4i\beta^{2}\int_{-\infty}^{\infty}dt''\int_{-\infty}^{0}dx''
G(x,t;x'',t'')G(x'',t'';x'',t'') \nonumber\\
G(x'',t'';x',t')\left(1+\frac{\epsilon^{2}}
{2(1+\sigma_{1})^{2}}e^{4x''}\right).
\end{eqnarray}

As it was shown in the previous section, the loop propagator  is
given by
\begin{eqnarray}
& &G(x'',t'';x'',t'')=i\int\int
\frac{d\omega''}{2\pi}\frac{dk''}{2\pi}\frac{1}{\omega''^{2}-k''^{2}-4+i\rho}
\left\{1+\frac{\epsilon^{2}}{4(1+\sigma_{1})^{2}}\,\frac{1}{(ik''+2)}e^{4x''}
\right.\nonumber\\ &
&\hspace{.6in}\left.-\frac{\epsilon^{2}}{4(1+\sigma_{1})^{2}}\,\frac{1}{
(ik''-2)}e^{4x''} \right.\nonumber\\ & &\hspace{.6in}\left.
+\left(K_{0}-\frac{\epsilon^{2}}{2(1+\sigma_{1})^{2}}\,\frac{1}{(ik''-2)}e^{4x''}
K_{0}+\epsilon K_{1}+\epsilon^{2} K_{2}\right)
e^{-2ik''x''}\right\}.
\end{eqnarray}
The above integral is logarithmically divergent.  Nevertheless,
the divergence of the loop integral can be removed by performing a
minimal subtraction.  Moreover,  doing the integration over
$\omega''$ then, we  obtain
\begin{eqnarray}
G(x'',t'';x'',t'')&=&\frac{\epsilon^{2}}{8\pi(1+\sigma_{1})^{2}}e^{4x''}
\nonumber\\ & &+\frac{1}{2}\int
\frac{dk''}{2\pi}\frac{1}{\sqrt{k''^{2}+4}}
\biggl(K_{0}-\frac{\epsilon^{2}}{2(1+\sigma_{1})^{2}}\,\frac{1}{(ik''-2)}e^{4x''}
K_{0} \biggr.\nonumber\\ & &\hspace{2in}\biggl.+\epsilon
K_{1}+\epsilon^{2} K_{2}\biggr) e^{-2ik''x''}.
\end{eqnarray}

Now regarding  the bulk contribution (9.1), the next step is to set
$k \rightarrow -k$ in the first term of the first propagator.
Secondly,  the integration over  $t''$ leads to the fact that the
energy variables of the first and the third propagators are equal.
Hence, the type I (bulk) contribution takes the following form:
\begin{eqnarray}
& &-4i\beta^{2}\int_{-\infty}^{0}dx''\int\int
\frac{d\omega}{2\pi}\frac{dk}{2\pi}\frac{ie^{-i\omega(t-t')}}{\omega^{2}-k^{2}-4+i\rho}
\Biggl\{\left(1+\frac{\epsilon^{2}}{4(1+\sigma_{1})^{2}}\,\frac{1}{(ik+2)}e^{4x}
\Biggr.\right.\nonumber\\ &
&\hspace{2.5in}\Biggl.\left.-\frac{\epsilon^{2}}{4(1+\sigma_{1})^{2}}\,\frac{1}
{(ik-2)}e^{4x''}\right) e^{-ik(x-x'')} \Biggr.\nonumber\\ &
&\hspace{.6in}\Biggl.
+\biggl(K_{0}-\frac{\epsilon^{2}}{4(1+\sigma_{1})^{2}}\,\frac{1}{(ik-2)}e^{4x}
K_{0}(k)-\frac{\epsilon^{2}}{4(1+\sigma_{1})^{2}}\,\frac{1}{(ik-2)}e^{4x''}
 K_{0}
\Biggr.\biggr.\nonumber\\ & &\hspace{3.5in}\Biggl.\biggl.+\epsilon
K_{1}+\epsilon^{2} K_{2}\biggr) e^{-ik(x+x')}\Biggr\} \nonumber\\
& &\times\,\Biggl\{\frac{\epsilon^{2}}{8\pi(1+\sigma_{1})^{2}}e^{4x''}
+\frac{1}{2}\int \frac{dk''}{2\pi}\frac{1}{\sqrt{k''^{2}+4}}
\biggl(K_{0}-\frac{\epsilon^{2}}{2(1+\sigma_{1})^{2}}\frac{1}{(ik''-2)}e^{4x''}
K_{0} \Biggr.\biggr.\nonumber\\ &
&\hspace{3.5in}\Biggl.\biggl.+\epsilon K_{1}+\epsilon^{2}
K_{2}\biggr) e^{-2ik''x''}\Biggr\} \nonumber\\ & &\times\,
\int\frac{dk'}{2\pi}\frac{i}{\omega^{2}-k'^{2}-4+i\rho} \Biggl\{
\left(1+\frac{\epsilon^{2}}{4(1+\sigma_{1})^{2}}\frac{1}{(ik'+2)}e^{4x''}
\Biggr.\right.\nonumber\\ &
&\hspace{2.5in}\Biggl.\left.-\frac{\epsilon^{2}}{4(1+\sigma_{1})^{2}}\frac{1}
{(ik'-2)}e^{4x'}\right) e^{ik'(x''-x')} \Biggr.\nonumber\\ &
&\hspace{.6in}\Biggl.
+\biggl(K_{0}-\frac{\epsilon^{2}}{4(1+\sigma_{1})^{2}}\frac{1}{(ik'-2)}e^{4x''}
K_{0}(k')-\frac{\epsilon^{2}}{4(1+\sigma_{1})^{2}}\frac{1}{(ik'-2)}e^{4x'}
 K_{0}
\Biggr.\biggr.\nonumber\\ & &\hspace{3in}\Biggl.\biggl.+\epsilon
K_{1}+\epsilon^{2} K_{2}\biggr) e^{-ik'(x''+x')}\Biggr\}
\nonumber\\ &
&\times\,\left\{1+\frac{1}{2}\frac{\epsilon^{2}}{(1+\sigma_{1})^{2}}e^{4x''}\right\}.
\end{eqnarray}
The  integration over $x''$ may be performed by the formula (5.9)
and therefore, (9.4) converts to
\begin{eqnarray}
& &-2\beta^{2}\int\int\int \frac{d\omega}{2\pi}\frac{dk}{2\pi}
\frac{dk'}{2\pi}
e^{-i\omega(t-t')}e^{-i(kx+k'x')}\frac{i}{\omega^{2}-k^{2}-4+i\rho}
\frac{i}{\omega^{2}-k'^{2}-4+i\rho} \nonumber\\ & &\int
\frac{dk''}{2\pi}\frac{\epsilon^{2}}{\sqrt{k''^{2}+4}}
\Biggl\{K_{0}(k'')\left(\frac{1}{4(1+\sigma_{1})^{2}}\frac{1}{ik+2}
\frac{1}{k+k'-2k''-4i} \Biggr.\right.\nonumber\\ &
&\hspace{2in}\Biggl.\left.+\frac{1}{4(1+\sigma_{1})^{2}}\frac{1}{ik'+2}
\frac{1}{k+k'-2k''-4i} \Biggr.\right.\nonumber\\ & &\Biggl.\left.
+\frac{1}{4(1+\sigma_{1})^{2}}\frac{1}{ik+2}
\frac{K_{0}(k')}{k-k'-2k''-4i}
+\frac{1}{4(1+\sigma_{1})^{2}}\frac{1}{ik'+2}
\frac{K_{0}(k)}{k'-k-2k''-4i}\Biggr.\right. \nonumber\\ &
&\Biggl.\left.-\frac{1}{4(1+\sigma_{1})^{2}}\frac{1}{ik'-2}
\frac{K_{0}(k')}{k-k'-2k''-4i}
-\frac{1}{4(1+\sigma_{1})^{2}}\frac{1}{ik-2}
\frac{K_{0}(k)}{k'-k-2k''-4i}\Biggr.\right. \nonumber\\ &
&\Biggl.\left. +\frac{1}{4(1+\sigma_{1})^{2}}\frac{1}{ik'-2}
\frac{K_{0}(k)K_{0}(k')}{k+k'+2k''+4i}
+\frac{1}{4(1+\sigma_{1})^{2}}\frac{1}{ik-2}
\frac{K_{0}(k)K_{0}(k')}{k'+k+2k''+4i}\Biggr.\right. \nonumber\\ &
&\Biggl.\left.
+\frac{1}{k-k'-2k''}K_{2}(k')+\frac{1}{k'-k-2k''}K_{2}(k)
\Biggr.\right.\nonumber\\ & &\Biggl.\left.
-\frac{1}{k+k'+2k''}K_{0}(k)K_{2}(k')-\frac{1}{k'+k+2k''}K_{0}(k')
K_{2}(k)\Biggr.\right.\nonumber\\ & &\Biggl.\left.\hspace{2in}
-\frac{1}{k+k'+2k''}K_{1}(k)K_{1}(k')\right)\Biggr. \nonumber\\ &
&\Biggl.+K_{1}(k'')\left(\frac{1}{k-k'-2k''}K_{1}(k')-
\frac{1}{k+k'+2k''}K_{0}(k)K_{1}(k')\right.\Biggr.\nonumber\\ &
&\Biggl.\left.\hspace{1in}+\frac{1}{k'-k-2k''}K_{1}(k)-
\frac{1}{k+k'+2k''}K_{0}(k')K_{1}(k)\right)\Biggr.\nonumber\\ &
&\Biggl.-\frac{1}{2(1+\sigma_{1})^{2}}\frac{K_{0}(k'')}{ik''-2}\left(
\frac{1}{k+k'-2k''-4i}+\frac{1}{k-k'-2k''-4i}K_{0}(k')\Biggr.\right.\nonumber\\
& &\Biggl.\left.\hspace{1in}+\frac{1}{k'-k-2k''-4i}K_{0}(k)-
\frac{1}{k+k'+2k''+4i}K_{0}(k)K_{0}(k') \right)\Biggr.\nonumber\\
&
&\Biggl.+K_{2}(k'')\left(\frac{1}{k+k'-2k''}+\frac{1}{k-k'-2k''}K_{0}(k')
\Biggr.\right.\nonumber\\ &
&\Biggl.\left.\hspace{1in}+\frac{1}{k'-k-2k''}K_{0}(k)-\frac{1}{k+k'+2k''}K_{0}(k)K_{0}(k')
\right)\Biggr.\nonumber\\ &
&\Biggl.+\frac{1}{2(1+\sigma_{1})^{2}}K_{0}(k'')\left(
\frac{1}{k+k'-2k''-4i}+\frac{1}{k-k'-2k''-4i}K_{0}(k')
\Biggr.\right.\nonumber\\ &
&\Biggl.\left.\hspace{.6in}+\frac{1}{k'-k-2k''-4i}K_{0}(k)
-\frac{1}{k+k'+2k''+4i}K_{0}(k)K_{0}(k') \right)\Biggr\}.
\end{eqnarray}
Meanwhile,  there is another term which is the remaining part of
the contribution i.e.:
\begin{eqnarray}
& & \frac{\beta^{2} \epsilon^{2}}{2\pi(1+\sigma_{1})^{2}}
\int\int\int \frac{d\omega}{2\pi}\frac{dk}{2\pi} \frac{dk'}{2\pi}
e^{-i\omega(t-t')}e^{-i(kx+k'x')}\frac{1}{\omega^{2}-k^{2}-4+i\rho}
\frac{1}{\omega^{2}-k'^{2}-4+i\rho} \nonumber\\ &
&\hspace{1in}\left( \frac{1}{k+k'-4i}+\frac{1}{k-k'-4i}K_{0}(k')
\right.\nonumber\\ &
&\left.\hspace{1.4in}+\frac{1}{k'-k-4i}K_{0}(k)
+\frac{1}{-k-k'-4i}K_{0}(k)K_{0}(k') \right).
\end{eqnarray}

Next, we need to integrate over  $k''$ however, the corresponding
calculations are tedious. We will discover two important facts
during the computations.  Firstly,  all the terms involving the
following kind of integral $$\int \frac{dk''}{\sqrt{k''^{2}+4}}
\frac{1}{k+k'-2k''}$$ which is proportional to $\theta$ (after
doing the  integrations over $k$ and $k'$ and using the fact that
$ \hat{k}=k=k'=2\sinh \theta$) will be cancelled.  This
 is  consistent for a comparison with Ghoshal's
formula. Secondly, there are many  other terms which contain this
type of integral $$\int \frac{dk''}{\sqrt{k''^{2}+4}}
\frac{1}{k+k'-2k''-4i}$$ and  all of them (after assembling) will
be  canceled  with the counterpart terms in the type III
(bulk-bulk) Feynman diagram as we mentioned in  section seven.
 This result is also compatible with Ghoshal's formula.
Meanwhile,  further simplification can be made by using the values
of $k$ and $k'$ given by their poles and simplifying the
integrand. So the type I (bulk) Feynman diagram has the following
contribution
\begin{eqnarray}
& &-\frac{i\beta^{2}\epsilon^{2}}{\pi}\int \frac{d\omega}{2\pi}
e^{-i\omega(t-t')}e^{-i\hat{k}(x+x')} \frac{1}{(2\hat{k})^{2}}
\nonumber\\ & &\left\{-\frac{\cos a_{1}\pi}{(1+\cos
a_{1}\pi)^{2}}\frac{a_{1}\pi} {\sin
a_{1}\pi}\left(\frac{1}{i\hat{k}+2}\,\frac{1}{i\hat{k}-2\cos a_{1}\pi+2}
+\frac{1}{i\hat{k}-2}\,\frac{K_{0}^{2}(\hat{k})}
{i\hat{k}+2\cos a_{1}\pi-2}\right)\right.\nonumber\\ &
&\left.
+\left(1-\frac{2a_{1}\pi\cos a_{1}\pi}{\sin
a_{1}\pi}\right)
\frac{1}{\sin ^{2}
a_{1}\pi}\frac{1}{\hat{k}^{2}+4}K_{0}(\hat{k})
 \right.\nonumber\\ &
&\left.+\left(-\frac{1}{4}+\frac{a_{1}\pi}{\sin a_{1}\pi}
\frac{2i\hat{k}}{i\hat{k}-2\cos a_{1}\pi}
+\frac{\pi}{\sqrt{\hat{k}^{2}+4}}\right)K_{2}(\hat{k})\right.
\nonumber\\ & &\left.-\left(\frac{1}{2\sin
^{2}a_{1}\pi}-\frac{a_{1}\pi \cos a_{1}\pi} {2\sin
^{3}a_{1}\pi}\right)\frac{2i\hat{k}}{i\hat{k}-2\cos
a_{1}\pi}K_{1}(\hat{k})+\frac{a_{1}\pi}{\sin a_{1}\pi}
\frac{i\hat{k}}{i\hat{k}+2\cos a_{1}\pi}K_{1}^{2}(\hat{k})\right.
\nonumber\\ & &\left.+\frac{a_{1}\pi\cos a_{1}\pi(2\cos
a_{1}\pi-3)} {2(1+\cos a_{1}\pi)\sin^{3}
a_{1}\pi}\left(
\frac{1}{i\hat{k}-2\cos a_{1}\pi+2}+\frac{K_{0}(\hat{k})}{1-\cos a_{1}\pi}+
\frac{K_{0}^{2}(\hat{k})}{i\hat{k}+2\cos a_{1}\pi-2}\right)
 \right. \nonumber\\ &
&\left.+\left(\frac{7}{12\sin^{2} a_{1}\pi}
+\frac{\cos a_{1}\pi}{2\sin^{4}
a_{1}\pi}\right)K_{0}(\hat{k})-\frac{2\cos a_{1}\pi}{\sin^{2} a_{1}\pi}\,
\frac{1}{(i\hat{k}-2\cos a_{1}\pi)^{2}}\right.
\nonumber\\ & &\left.+\frac{1}{4}\left(\frac{a_{1}\pi}{\sin
a_{1}\pi}-1\right) \frac{\cos a_{1}\pi}{\sin^{4}
a_{1}\pi}\,\frac{4\cos^{2}a_{1}\pi-
\hat{k}^{2}}{(i\hat{k}-2\cos a_{1}\pi)^{2}} \right.
\nonumber\\ & &\left.-\frac{a_{1}\pi}{2\sin a_{1}\pi}
\left(K_{2}(\hat{k})+K_{0}^{2}(\hat{k})K_{2}(-\hat{k})
\right)\right. \nonumber\\ & &\left.
+\frac{1}{2\sin^{4}a_{1}\pi}\left(2\cos a_{1}\pi
-\frac{a_{1}\pi\left(1+3\cos^{2}a_{1}\pi\right)}
{2\sin a_{1}\pi}\right) \,
\frac{\hat{k} ^{2}}{(i\hat{k}-2\cos a_{1}\pi)^{2}}
\right\}.
\end{eqnarray}

\resection{ Ghoshal's formula up to the second order}

 Ghoshal's formula (2.8) for the sinh-Gordon quantum reflection factor up to one
loop order is given by:
\begin{equation}
K_{q}(\theta) \sim K_{c}(\theta) \left( 1-\frac{i \beta^{2}}{8}
\sinh \theta\ {\cal F}(\theta) \right),
\end{equation}
where
\begin{eqnarray}
{\cal F}(\theta)&=& \frac{1}{\cosh\theta+1} -
\frac{1}{\cosh\theta}\nonumber \\ & &+
\frac{e_{1}}{\cosh\theta+\sin(e_{0}\pi/2)}
-\frac{e_{1}}{\cosh\theta - \sin(e_{0}\pi/2)} \nonumber \\ & &+
\frac{f_{1}}{\cosh\theta+\sin(f_{0}\pi/2)}
-\frac{f_{1}}{\cosh\theta-\sin(f_{0}\pi/2)}.
\end{eqnarray}
In calculating (10.2) we have made use of the expansions of $E$ and
$F$ to $O(\beta^2)$:
\begin{equation}
E\sim e_0 +e_1{\beta^2\over 4\pi} \qquad F\sim f_0 +
f_1{\beta^2\over4\pi},
\end{equation}
where
\begin{equation}
e_{0}=a_{0}+a_{1} \hspace{.5in} \hbox{and} \hspace{.5in}
f_{0}=a_{0}-a_{1}.
\end{equation}
Since $K_q=K_c+\delta K_c$, we find that
\begin{equation}
\frac{\delta K_{c}}{K_{c}}=-\frac{i\beta^{2}}{8}\sinh \theta\
{\cal F}(\theta).
\end{equation}
Hence, expanding to $O(\epsilon^{2})$, we obtain
\begin{eqnarray}
{\cal F}(\theta)&=&\Biggl\{
\frac{1}{\cosh\theta+1}-\frac{1}{\cosh\theta} +
\frac{e_{1}}{\cosh\theta+\sin a_{1}\pi}-
\frac{e_{1}}{\cosh\theta-\sin a_{1}\pi} \Biggr.\nonumber \\ &
&\hspace{.1in}\Biggl.+ \frac{e_{1}\epsilon \cos a_{1}\pi}{2\sin
a_{1}\pi}\left( \frac{1}{(\cosh\theta+\sin a_{1}\pi)^{2}} +
\frac{1}{(\cosh\theta-\sin a_{1}\pi)^{2}}
\right)\Biggr.\nonumber\\& &
\hspace{.1in}\Biggl.+\frac{e_{1}\epsilon^{2} }{8\sin
a_{1}\pi}\left( \frac{1}{(\cosh\theta+\sin a_{1}\pi)^{2}} +
\frac{1}{(\cosh\theta-\sin a_{1}\pi)^{2}}\right)\Biggr.\nonumber\\
& & \hspace{.1in}\Biggl.+\frac{e_{1}\epsilon^{2}\cos^{2}a_{1}\pi
}{4\sin^{2} a_{1}\pi}\left( \frac{1}{(\cosh\theta+\sin
a_{1}\pi)^{3}} - \frac{1}{(\cosh\theta-\sin
a_{1}\pi)^{3}}\right)\Biggr.\nonumber\\
 & &\hspace{3in}\Biggl.+\frac{\epsilon
f_{1}}{\sin a_{1}\pi \cosh^{2}\theta}\Biggr\},
\end{eqnarray}
where \cite{CC}
\begin{equation}
e_{1}=-2a_{1}+ \frac{\epsilon}{\pi \sin a_{1}\pi} +O(\epsilon^2)
\end{equation}
and $f_{1}$ is proportional to $\epsilon$.

On the other hand,  the relative correction to the classical
reflection factor $K_{c}$ is given by (using (3.20))
\begin{eqnarray}
& &\frac{\delta K_{c}}{K_{c}}=K_{0}^{-1}\delta K_{0} +\epsilon
\left(K_{0}^{-1} \delta K_{1}- K_{0}^{-2}K_{1} \delta
K_{0}\right)\nonumber\\
& & \hspace{.5in}+\, \epsilon^{2} \left(K_{0}^{-1}  \delta
K_{2}-K_{0}^{-2} K_{1} \delta K_{1} -K_{0}^{-2} K_{2} \delta K_{0}
+K_{0}^{-3} K_{1}^{2} \delta K_{0}\right).
\end{eqnarray}
In fact, the corrections to $K_{0} (\delta K_{0})$ and $K_{1} (\delta K_{1})$
have been calculated in Ref. \cite{C} and Ref. \cite{CC},
respectively. Clearly, all the contributions that have been
obtained in this paper are involved in the correction to $K_{2}$
i.e. $\delta K_{2}$.  It is seen  that the functional form of the
second order quantum corrections to the classical reflection factor
corresponding to the type I and type III Feynman diagrams are
consistent with Ghoshal's formula.

\resection{Discussion}
In this paper we studied the sinh-Gordon model restricted to a
half-line by boundary conditions which are compatible with
integrability. We calculated the second order quantum corrections
to the classical reflection factor of the model. The principal
purpose of the perturbative calculation is to test the quantum
reflection factor of the sinh-Gordon particle given by Ghoshal up
to the second order in the difference of the two boundary
parameters. In fact, Ghoshal found the general form of the quantum
reflection factor of the sinh-Gordon model. However, apart from
two special cases (Neumann and Dirichlet boundary conditions)
Ghoshal's formula could not provide a complete relationship
between the reflection factor and the boundary data. As we
mentioned before, on the basis of a perturbative calculation, the
relationship between the various parameters was conjectured to be
\cite{CC}

\begin{equation}
E=(a_0+a_1)(1-B/2), \qquad F=(a_0-a_1)(1-B/2).
\end{equation}
Corrigan et.al \cite{CT} tried to give further support for these
formulae. They used two different technique i.e. the boundary
bootstrap method and the WKB semi-classical approximation
approach. Their non-perturbative calculations have been performed
for the complicated  case of the general boundary conditions in which
the bulk symmetry is violated. Alternatively, the conjecture (11.1)
may be checked on the basis of perturbative calculations up to the
second order. It is understood that this paper provides most part
of the collection of ingredients one need.

The computations associated with the type II Feynman diagram have
not been performed in this article. In fact, in this diagram the
middle momenta will be related to each other in an intricate
manner due to the double Green functions. Nevertheless, it is
expected that the same procedure will be applicable to the
remaining diagram as well. Moreover, many formulae of this paper
will be helpful.

All the affine Toda field theories whose corresponding  Lie algebras
are simply laced or non-simply laced algebras have known exact
quantum S-matrices. However, in the presence of a boundary the
boundary S-matrices or the reflection factors of the theories are
largely unknown. In other words, the analogue of Ghoshal's formula
has not been formulated for most of the models. Besides,
classifying the quantum integrability of the affine Toda field
theories has not been completed.

\resection{Acknowledgment}
The author wishes to thank E. Corrigan and P. Bowcock for
discussions and fruitful comments.

\end{document}